# Preventing Disclosure of Sensitive Knowledge by Hiding Inference

A.S.Syed Navaz,          M.Ravi,          T.Prabhu

[123]Department of Computer Applications,

Muthayammal College Of Arts & Science, Namakkal, India.

## ABSTRACT

'Data Mining' is a way of extracting data or uncovering hidden patterns of information from databases. So, there is a need to prevent the "inference rules" from being disclosed such that the more secure data sets cannot be identified from non-sensitive attributes. This can be done through removing/adding certain item sets in the transactions (Sanitization). The purpose is to hide the Inference rules, so that the user may not be able to discover any valuable information from other non-sensitive data and any organisation can release all samples of their data without the fear of 'Knowledge Discovery In Databases' which can be achieved by investigating frequently occurring item sets, rules that can be mined from them with the objective of hiding them. Another way is to release only limited samples in the new database so that there is no information loss and it also satisfies the legitimate needs of the users.

The major problem is uncovering hidden patterns, which causes a threat to the database security. Sensitive data are inferred from non-sensitive data based on the semantics of the application the user has, commonly known as the 'inference problem'. Two fundamental approaches to protect sensitive rules from disclosure are that, preventing rules from being generated by hiding the frequent sets of data items and reducing the importance of the rules by setting their confidence below a user-specified threshold.

## 1. INTRODUCTION

Organisations collect data in orders of magnitude greater than ever before. Data mining techniques such as Classification mining, Association rules, Functional dependency are available for efficient analysis of sensitive data. These techniques detect relationships or associations between specific values of categorical variables in large data sets. This is a common task in many data mining projects. These powerful exploratory techniques have a wide range of applications in many areas of business practice and also research. These techniques enable analysts and researchers to uncover hidden patterns in large data sets. Sensitive information must be protected against unauthorized access. The protection of the confidential information has been a long-term goal and the project suggests a method of hiding the inference rules and thereby prevents the disclosure of sensitive information.

### 1.1 Problem Definition

The major problem incurred in databases, using data mining techniques is that, they pose a threat to 'Database Security'.[1],[2]. Inference problem occurs because of uncovering hidden patterns of information from the database using non-sensitive data. Inference problem refers to finding out or inferring data that is more sensitive. Association Rules are a useful technique for extracting data from databases. Even though, the Association Rules give only a probabilistic knowledge of the data, it may be an advantage for the competitor to gain some interesting knowledge from the databases. Sensitive data are inferred from non-sensitive data based on semantics of the application the user has. The data may require some information, but certain data are not to be disclosed. There should not be information loss in the database. Hence, there is also a need for confidentiality compromise between disclosed information and required needs of the data users.

### 1.2 State of Art

Some algorithms already exist for the proposed problem[1]. The first algorithm hides the sensitive rules by increasing the support of the rule until the confidence decreases below the minimum confidence threshold. The next algorithm decreases the support of the sensitive rules until either their confidence is below the minimum confidence threshold or their support is below the minimum support threshold. The hiding strategies that were used heavily depend on finding transactions that fully or partially support the generating item sets of a rule. Another issue is that the changes in the database introduced by the hiding process should be limited in such a way that the information loss incurred by the process is minimal. The proposed algorithms try to apply only minimal changes in the database at every step of the hiding process.

But certain assumptions were made in the algorithms like:

- The algorithms hide disjoint rules
- The rules were hidden at one time

#### *1.2.1. New Rules:*

New rules are those rules that could not be mined in the source database, but can be retrieved after the hiding process.

#### *1.2.2. Lost Rules:* Lost rules are those rules that can be mined from the source database and cannot be retrieved in the released database.

The time complexity increases due to exhaustive search of frequent item sets which is obviously infeasible, except for small item sets. The source database gets augmented and fuzzified due to the changes made.

Overlapping rules were not considered in the existing algorithms. When hiding an item in a particular rule, there is the need to haunt back and front, if the overlapping rules were taken into consideration. Hiding an item, also forces to hide some other item and the time complexity increases. This was a major disadvantage in the existing system

## 1.3. Block diagram

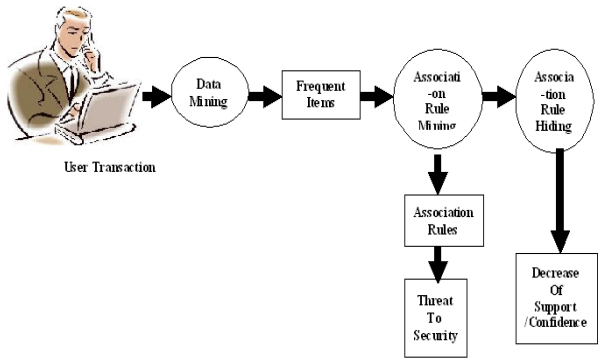

**Fig 1 Data Mining**

The proposed method aims at preventing disclosure of sensitive data through the combination of inference rule of non-sensitive data. The proposed method has two approaches in order to protect sensitive rules from disclosure [3]. The first one is to prevent the rules from being generated by hiding the frequent sets of data items from which they are derived and also minimise the search time for the items The second is to reduce the importance of the rules by setting their confidence below a user-specified threshold, so that no rules can be derived from the selected items.

Figure 1. Represents the overall block diagram the proposed method. These approaches also enhance the speed and see to that there is no information loss & no additional information added. All the assumptions in the existing system are dropped such as the overlapping rules are hidden, sensitive data are considered and many rules are hidden at a time.

## 2. DATA MINING

Data mining techniques are used to identify the frequent item sets, the transactions and all sets of user interactions from the database. Association rule helps to achieve the purpose. Through association rule mining, all possible rules can be extracted from the existing database. Possible solutions to prevent data mining technique from releasing the source database are augmenting the database and fuzzifying the database. But preventing the Data Mining techniques from disclosing the data should be done with the major constraint that there is less information loss, less distortion and no trace of widening the database.

### 2.1. Fundamentals of Data Mining

Data Mining also called as Knowledge Discovery in Databases (KDD) is the process of extracting useful information and patterns from the database[2]. It predicts Data Mining (DM) techniques and related applications have however increased the security risks that one may incur when releasing data. The elicitation of knowledge that can be attained by such techniques has been the focus of the Knowledge Discovery in Databases.

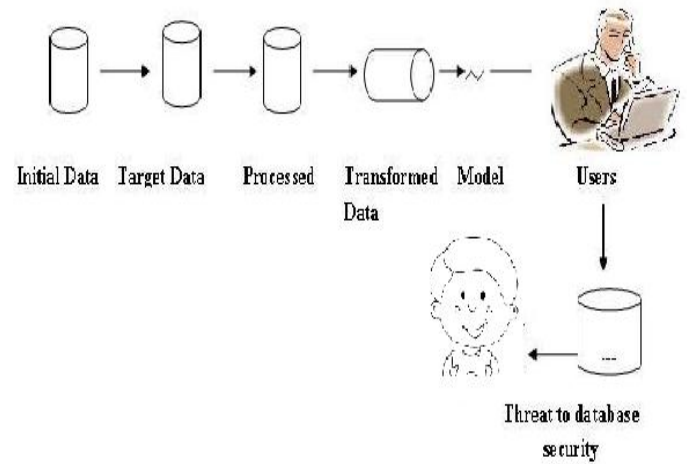

**Fig 2.Process of data mining**

### 2.2. Need for Data Mining

The major reason that data mining has took a great attention in information technology is due to the wide availability of huge amount of data and the need for turning that data into useful information and knowledge. The information can be used for applications ranging from Market-Basket analysis up to production control. Data from databases remain as archives that can be visited by everyone. Consequently, important decisions are often made based on the sensitive information in the databases. Though a probabilistic knowledge could only be gained from the Data Mining process, there is an immediate need to prevent the disclosure of the sensitive information.

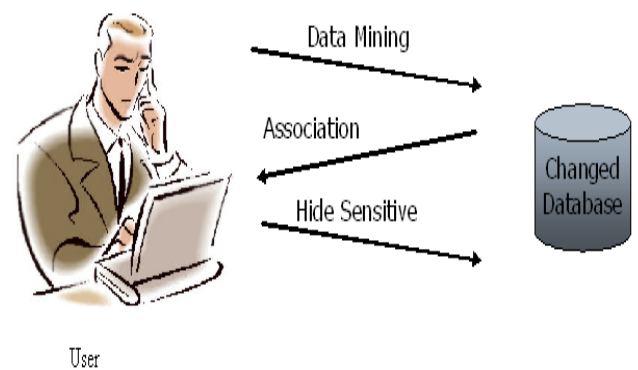

**Fig 3. Data Mining and Inference Rule Hiding**

Figure 3 refers to Association Rule hiding in the Data Mining techniques, for the purpose of hiding the sensitive information [4]. The user uses various Data Mining Techniques like Inference Rules, Functional dependencies to extract information from the database. Association Rules help to retrieve the sensitive data from non-sensitive data. There is a need for any business organizations, to hide these rules such that the competitors are prevented from viewing the original database.

## 3. INFERENCE RULE HIDING

Information systems contain confidential information such as social security number, income, type of disease, weapons etc., that must be properly protected. Existing techniques like deduction rules, Association rules, functional dependency discloses sensitive data, which is commonly known as 'Inference problem'. Data Mining helps to retrieve the interesting association rules. Inference Rule Hiding aims to hide these Association rules so that no valuable information can be mined from the database.

The steps involved are:

- Identifying frequently occurring item sets
- Finding all the transactions that support the item sets
- Retrieve all the possible Association rules and
- Finally hide these association rules by decreasing their support/confidence.

### 3.1. Basic Concepts

Association rules referred to as link analysis is one of the techniques in data mining. These associations are often used in the retail sales community to identify items that are frequently purchased together. They are of the form, $X \Rightarrow Y$ (the purchasing of the product Y depends on the purchase of X). The efficiency of association rule algorithms depends on the number of scans of the database that are required to find out the frequently occurring item sets. The support of an item in an association rule is the percentage of transactions in which that item occurs. The confidence or strength for an association rule $X \Rightarrow Y$ is the ratio of the number of transactions that contain $X \cup Y$ to the number of transactions that contains X. Hence, the project aims to achieve the objective of hiding the association rules by decreasing the support of items and the confidence of the rules.

### 3.2. Module Description

The project consists of three phases namely:
- Finding maximum frequent items
- Association rule mining
- Association rule hiding

Each and every phase uses a new algorithm that has not yet been explored.

### 3.2.1. Finding maximum frequent items

The first module uses the "Apriori Algorithm" to search for the maximum frequent items occurring in the selected database. The number of large itemsets depends on the user specified support and confidence. The search for the large itemsets is based on the concept of searching using passes where only the candidate itemsets are passed on to subsequent passes.

### 3.2.1. Association rule mining

The second module uses the standard rule mining algorithm to mine the rules from the commonly occurring itemsets. The rules mined take into account the entire singleton and also the subsets of itemsets to mine the rules. From the confidence threshold value, entered by the user, the set of all possible Association Rules are mined. The support value is considered for extracting the Maximum Frequent Itemset, whereas confidence value is a must for mining the rules.

### 3.2.2. Association rule hiding

The third module uses the "Weight Based Sorting Distortion Algorithm" to hide the items that are frequently occurring and also the valuable inference rules. The algorithm considers both the disjoint and the joint rules of interest and decreases their support/confidence below a user-specified threshold to hide the rules. The user can compare the original database with the sanitized database, after hiding. Both the Large Itemset and the Association Rules are reduced, based on the weight of the items calculated.

### 3.3. Maximum Frequent Sets

The database contains large amount of items, in which certain items occur more frequently. The frequently occurring itemsets are the source for mining the Association Rules. An item is said to be a large itemset when its number of transactions (support) is above a minimum support threshold. According to the support value the Large Itemset also varies. When the large itemsets are found out, the rules can be easily mined from the confidence and support threshold values. The Large Itemset serves as an input to find Association Rules. Any interesting Association Rules of the form $X \Rightarrow Y$ can be mined from the Maximum Frequent Itemsets.

The approach used to count the large itemset is quite simple. Given a set of items of size m, there are 2m subsets. Since we are not interested in the empty set, the potential number of large itemsets is 2m-1. The potentially large itemsets are called 'candidates'. The performance measure used for association rule algorithm is the size of the candidates. The number of scans used to search for the Maximum Frequent Itemsets is less, when using the Apriori Algorithm. In the first pass the singleton items are scanned and for the second, third, fourth etc., up to nth passes they are combined in pairs of two, three, etc., to generate subsets. After completing the required scans, the item which has the maximum support could be easily identified. Every Large Itemset is combined with every other singleton item, to generate subsets. For a support of 100%, there is less chance to find Maximum Frequent Itemsets because, it is impossible for all the items to occur in all the transactions. If the minimum support is 0%, all subsets of item will form the Maximum Frequent Itemsets.

### 3.3.1 Apriori Algorithm

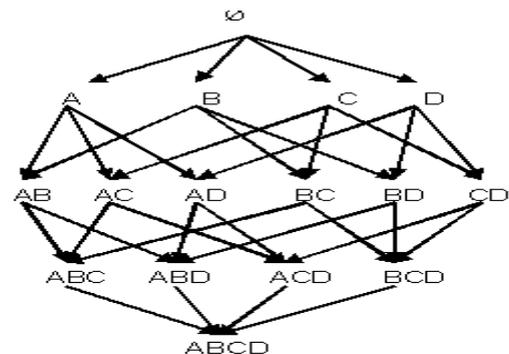

**Fig 4. Subsets of Item sets in Apriori**

Figure 4.represents the formation of subsets using the Apriori Algorithm.The basic idea behind the algorithm is to generate the candidate item of a particular size and then scan the database to count them to see if they are large. Only those candidates that are large are used to generate candidates for the next pass. An itemset is considered as a candidate only if all its subsets are also large. All singleton items are used as candidates in the first pass. After the first scan every large itemset is combined with every other large item. The maximum number of database scans is one more than the cardinality of the largest itemset.

### 3.3.2 Apriori Algorithm-Example

The database 'D' contains Transaction ID's and Itemsets. C1 is used as candidate in the first pass. Singleton items are the only inputs to the first pass. The minimum support threshold set by the user is 2. So, only the candidates that have a support greater than 2 are called Large Itemsets (denoted as L1).

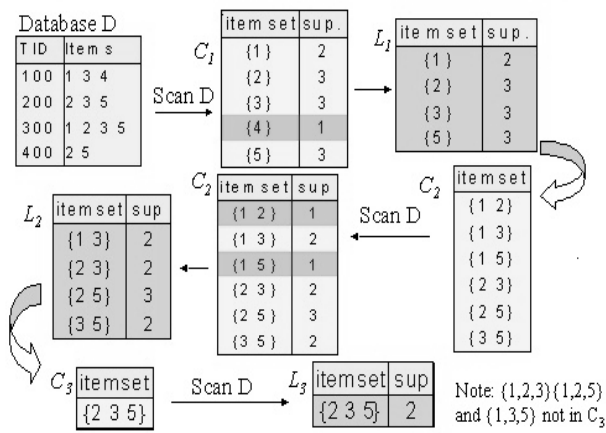

**Fig 5. Apriori example**

In the next pass, pairs of subsets are considered. Again the subsets that have a support of 2 or greater than 2 are the valuable inputs to the next scan. Finally L1, L2, L3 are the large Itemsets. The number of passes or scans required is 1 more than the cardinality of the largest itemset. The steps followed in the Apriori Algorithm are shown in Figure 5.

## 4. ASSOCIATION RULE MINING

Association rule mining is a two-step process: Find all frequent itemsets. By definition, each of these itemsets will occur at least as frequently as a predetermined minimum support count, generate strong association rules from the frequent itemsets. By definition, these rules must satisfy minimum support and minimum confidence.

A typical example of association rule mining is market basket analysis. This process analyses the customer buying habits by finding associations between the different items that customers place in their shopping basket. The discovery of such associations helps competitors to gain insight into which items are frequently purchased together by the customers. For instance, if the customers are buying milk, how likely are they to buy bread on the same time of purchase?

### 4.1 Association Rule Mining Algorithm

For every large itemset, find their subsets. Consider all the subsets in combinations of one, two, three, etc., to generate valid rules. If a particular rule does not has confidence, greater than the threshold there is no need to check for its subsets. With respect to the support and confidence, the large itemsets are identified. From the confidence threshold, rules are formed. The major statistics computed for the association rules are Support (relative frequency of the of the rule), Confidence (conditional probability of the rule) and Correlation between them.

The rules are mined according to the user-entered value for confidence. Any rule that has a confidence greater than the entered value, is considered to be valuable. The Association Rules are a valid input to the next module. From the set of Association Rules, some sensitive Rules are found and then the process of hiding the rules is carried out.

## 5. ASSOCIATION RULE HIDING

The hiding method includes reducing the support of frequent itemsets containing sensitive rules, reduce the confidence or support of rules. Decreasing confidence of rule involves increasing the support of X in transactions not supporting Y and decreasing the support of Y in transactions supporting both X and Y. Decreasing support of rule involves decreasing the support of the corresponding large itemset. There are two strategies for hiding the association rules namely Blocking and distortion.

The blocking-based technique aims to put a relatively small number of uncertainties and reduce the confidence of sensitive rules, But, the problems were: an adversary can easily infer the hidden values if he applies a smart inference technique. This can be overcome by inserting many uncertainties, but the process becomes complicated like, both 0's and 1's must be hidden, because if only 1's were hidden the adversary would simply replace all the uncertainties with 1's and would restore easily the initial database

The distortion-based technique is simple and the uncertainty is reduced when compared to blocking based techniques. In distortion-based technique, the itemsets that have a support less than the minimum support threshold are eliminated. The rules that have a confidence less than the minimum confidence threshold are also eliminated. In this way, by eliminating certain itemsets and some rules, the original database gets distorted. A comparison shows that the source database and the released database have differences in the cardinality of the large itemsets and the rules that are mined from them. Some of the challenges of distortion-based technique include: To minimize the undesirable Side Effects that the hiding process causes to non-sensitive rules. To minimize the number of 1's that must be deleted in the database. Algorithms must be linear in time as the database increases in size.

## 5.1. Weight Based Sorting Distortion

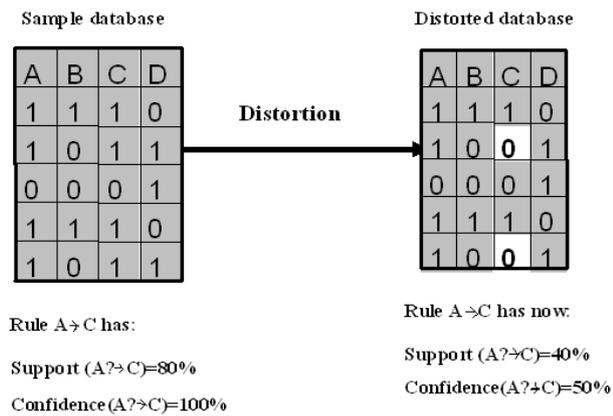

**Fig 6.Distortion Based Techniques**

Figure 6 displays the use of Distortion Technique on the sample database. After mining the association rules, the Weight based Sorting Distortion algorithm[5], [6] performs the following steps:

- Retrieve the set of transactions that support sensitive rules.
- For each rule find the number of transaction in which, one item will be hidden.
- For each rule in the database with common items, compute a weight that denotes how strong the rule is.
- For each transaction compute, how many strong rules this supports.
- Sort the transactions in ascending order according to the itemsets.
- For each transaction hide an item that is contained in the rule
- Update confidence and support values for the rules in the database.

## 6. IMPLEMENTATION

## 6.1. Maximum Frequent Item sets

### 6.1.1. Apriori Algorithm

**Input**

**I         //Itemsets**

**D         //Database of transactions**

**S         //Support**

**Output**

**L         //Large itemsets**

K=0;      //K is used as the scan number.

L=0;

C1=I;    //Initial candidates are set to be the items.

Repeat

K=k+1;

Lk=0;

for each Ii ε Ck do

Ci=0;       //Initial counts for each itemset are 0.

For each tjε D do

For each Iiε Ck do

if Iiεtj then

Ci=ci+1;

for each Ii ε Ck do

if ci>(sx \D|) do

Lk=L∪kIi;

L=L∪Lk;

Ck+1=Apriori-Gen (Lk)

Until Ck+1=0;

### 6.1.2 Apriori-Gen Algorithm

**Input**

**Li-1   //Large itemsets of size i-1**

**Output**

**Ci    //Candidates of size i**

Ci=0;

For each I ε Li-1 do

For each J <> I ε Li-1 do

If i-2 of the elements in I and J are equal then

Ck=Ck∪ {I∪J};

## 6.2. Association Rule Mining

**Input**

**D           //Database of transactions**

**I         //Items**

**L          //Large Itemsets**

**S          //Support**

**α         //Confidence**

**Output**

**R          //Association Rules satisfying s and α**

R=0;

For each I ε L do

For each X ε I do

If   support (I) >= α then

    support (x)

R=R U {x ⇒ (i-x)};

## 6.3. Association Rule Hiding Weight Based Sorting Distortion Algorithm

**Input**

**D        //Database**

**SR       //Sensitive Rules**

**AR** //Association Rules

**SM** //Safety Margin

**Priority** //stores sorted transaction

**N** //Transactions

**SDB** //Sanitized Database

**Output**

**DB** //Sanitized database

**Res** //result

### 6.3.1 Sort the database in ascending order

For t = 0 To Itemsets

For j = t + 1 To Itemsets - 1

If Priority (t) > Priority (j) Then

Priority (t) = Priority (j)

Priority (j) = t1

End If

Next

Next

For t = 0 Itemsets to - 1

Res = Res + Itemset + Priority (t)

Next

### 6.3.2. Hide and Distort the Database

For t = 0 To N1 - 1

For j = 0 To 0

tItem = SR.getAssociationRule(j).getX().getItem(0)

For k = 0 To SDB. Itemsets - 1

If SDB.Itemset(k).getEqualCount Then

Index = k

Exit For

End If

Next

If tIndex <> -1 Then

SDB.Itemset(tIndex).removeItem(tItem)

End If

TRS.Itemset(t).removeItem(tItem)

 Next

 Next

 Res =  Res + SDB

## 7. COMPARISON OF PERFORMANCE

The performance of the new algorithms when compared to the existing approaches is simple and does not require much time for implementation. The side effects in terms of new rules and lost rules are also minimised. Using less number of scans in the Apriori algorithm reduces the search time for the frequent itemsets in the first module. In the second module, the Association Rules are mined for the overlapping itemsets which were not considered previously. WSDA used for the third module is a very efficient hiding strategy that is based on distortion-based technique.

## 8. CONCLUSION

Discovering the maximum frequent sets and mining the association rules from the database are a threat to the security. It is very simple for anyone to discover the maximum frequent sets using algorithms apart from Apriori algorithm. Once the commonly occurring items have been discovered, mining the association rules is a simple task. The project aims to reduce the time taken for searching the frequent itemsets and also hides the interesting patterns of rules that can be mined. In general, the sensitive patterns of information are prevented from disclosure by a simple hiding strategy, of the valid association rules[7]. Concluded by stating the high percentage of security this could be achieved by following the Association Rule Hiding. The advantages of the proposed method are:

- Increased database security
- Less time for hiding process
- Time complexity is reduced for searching the more frequent items
- Overlapping itemsets are also considered

## 9. FUTURE ENHANCEMENT

The future enhancement of the project is that, the impact of other Data Mining techniques like classification mining, clustering, etc., on database security can be considered. Further reduce the confidence/support of the rules to the minimum value possible so that no rules can be mined. The performance of the algorithm can be improved by considering the side effects and even more reducing the time complexity.

## 11. Author's Profile

**A.S.Syed Navaz** received BBA from Annamalai University, Chidambaram 2006, M.Sc Information Technology from KSR College of Technology, Anna University Coimbatore 2009, M.Phil in Computer Science from Prist University, Thanjavur 2010 and M.C.A from Periyar University, Salem 2010 .Currently he is working as an Asst.Professor in the Department of Computer Applications, Muthayammal College of Arts & Science, Namakkal. His area of interests are Wireless Networks and Mobile Communications.

**M.Ravi** received B.Sc Mathematics from Thiruvalluvar Govt. Arts College. Madras University 2005, M.Sc Computer Science from Muthayammal College of Arts & Science, Periyar University 2008, M.Phil in Computer Science from Prist University, Thanjavur 2010 and B.Ed from Tamilnadu Teaching Education 2012. Currently he is working as an Asst.Professor in the Department of Computer Applications, Muthayammal College of Arts & Science, Namakkal. His area of interests are Computer Networks and Mobile Communications.

**T.Prabhu** received B.Sc Computer Science in Muthayammal College of Arts & Science, Periyar University 2005, & M.C.A., from Institute of Road & Transport Technology, Anna University Chennai 2008. Currently he is working as an Asst.Professor in the Department of Computer Applications, Muthayammal College of Arts & Science, Namakkal. His area of interests are Computer Networks and Mobile Communications.